\documentclass[letterpaper, 10 pt, conference]{ieeeconf}  

\IEEEoverridecommandlockouts                              
\usepackage{amsmath}
\usepackage{amssymb}
\usepackage{mathtools}
\usepackage{algpseudocode}
\usepackage[dvipsnames]{xcolor}
\usepackage{setspace}
\usepackage{multirow}

\newtheorem{thm}{Theorem}
\newtheorem{lem}{Lemma}

\newtheorem{defn}{Definition}
\newtheorem{rem}{Remark}

\title{\LARGE \bf
An Efficient Off-Policy Reinforcement Learning Algorithm for the Continuous-Time LQR Problem
}

\author{Victor G. Lopez and Matthias A. Müller
\thanks{This work received funding from the European Research Council (ERC) under the European Union’s Horizon 2020 research and innovation programme (grant agreement No 948679).}
\thanks{V. G. Lopez and M. A. Müller are with the Leibniz University Hannover, Institute of Automatic Control, 30167 Hannover, Germany
        {\tt\small \{lopez,mueller\}@irt.uni-hannover.de}}%
}

\begin{document}

\maketitle
\thispagestyle{empty}
\pagestyle{empty}

\begin{abstract}

In this paper, an off-policy reinforcement learning algorithm is designed to solve the continuous-time LQR problem using only input-state data measured from the system. Different from other algorithms in the literature, we propose the use of a specific persistently exciting input as the exploration signal during the data collection step. We then show that, using this persistently excited data, the solution of the matrix equation in our algorithm is guaranteed to exist and to be unique at every iteration. Convergence of the algorithm to the optimal control input is also proven. Moreover, we formulate the policy evaluation step as the solution of a Sylvester-transpose equation, which increases the efficiency of its solution. Finally, a method to determine a stabilizing policy to initialize the algorithm using only measured data is proposed.

\end{abstract}

\section{INTRODUCTION}

Reinforcement learning (RL) is a set of iterative algorithms that allow a system to learn its optimal behavior as it interacts with its environment \cite{SuBa:18,Bert05}. In the context of linear optimal control, RL has been used in the last few decades to solve the linear quadratic regulator (LQR) problem in continuous-time \cite{VrPaAbLe:09,Vamv:17,MoSoJo:20,LePaCh:12,PoSa:22,JiJi:12} and in discrete time \cite{BrYdBa:94,FaGeShMe:18,LopezAlsMue2023,YaKiMoXu:23}. For applications of RL procedures to nonlinear systems and other extensions, the reader is referred to the surveys \cite{KiVaMoLe:18,JiBiGa:20,Liuetal:21} and the references therein.

In the continuous-time linear time-invariant (CT-LTI) case, several RL algorithms with attractive properties have been designed. Although the first proposed algorithms required at least partial knowledge of the system model (e.g., \cite{VrPaAbLe:09}), completely data-based methods are now well known \cite{Vamv:17,MoSoJo:20,LePaCh:12,PoSa:22}. These data-based algorithms replace the need for model knowledge by measuring persistently excited data directly from the system. Most of these data-based methods are \emph{on-policy} algorithms, meaning that they require the application (or simulation) of an exciting input to the system at every iteration, such that a new set of data can be collected. In contrast, the authors in \cite{JiJi:12} proposed a data-based \emph{off-policy} RL algorithm. This method has the advantage of requiring to collect data from the system only once, and then every iteration of the algorithm is performed using the same batch of measurements. 

The method in \cite{JiJi:12}, as well as most on-policy methods, is formulated as the problem of determining the values of certain unknown matrices from a set of equations derived from the Bellman equation. Taking advantage of the properties of the Kronecker product, this problem is then expressed as a set of linear equations that can be easily solved. However, the Kronecker product formulation generates matrices of large size, and this procedure presents a high computational burden that increases rapidly with an increase in the system dimension.

Another important issue in the existing learning-based control literature is the selection of a proper persistently exciting (PE) input. In most of the above literature, heuristic approaches for persistence of excitation are employed, often designing exciting inputs by adding sinusoidal, exponential and/or random signals \cite{JiBiGa:20}. A different approach for persistence of excitation was studied in \cite{WillemsRapMarDe2005}, where conditions for the design of a discrete-time PE input are formally established. It is shown in \cite{WillemsRapMarDe2005} that their definition of persistence of excitation provides data measurements that are so rich in information that every possible trajectory of a controllable discrete-time linear system can be expressed in terms of such data. This result is now known as Willems' lemma, and has been successfully used in recent years in data-based analysis, estimation and control of discrete-time systems (see, e.g., the survey \cite{MarkovskyDor2021} and the references therein). In \cite{LePaCh:12}, it was proposed to use a PE signal as defined in \cite{WillemsRapMarDe2005} to excite a continuous-time system during a Q-learning procedure, which guarantees solvability of their policy evaluation step. However, the method in \cite{LePaCh:12} is an \emph{on-policy} algorithm and the authors require persistence of excitation of a signal composed of both the input and the state of the system. This contrasts with our objective of considering a PE signal in terms of the input only. Moreover, in \cite{LePaCh:12} a high order of persistence of excitation is needed.

The contributions of this paper are as follows. We propose a novel data-based \emph{off-policy} RL algorithm to solve the LQR problem for continuous-time systems. As in \cite{JiJi:12}, we perform the policy evaluation and policy improvement steps simultaneously. Different from the existing algorithms, we formulate a Sylvester-transpose equation that can be efficiently solved using known methods \cite{Hajarian2016,TansriChoCha2022,SongCheZha2011}. This avoids the use of the Kronecker product and the ensuing large matrices in our computations. Moreover, we use the results in \cite{LoMu:22}, where a continuous-time version of Willems' lemma was proposed. This allows us to design a PE input that guarantees the solvability of the Sylvester-transpose equation in a data-based fashion. In our formulation, persistence of excitation depends only on the input of the system, and we require the use of a PE input of lower order compared to \cite{LePaCh:12}. Finally, we propose a method to determine the required initial stabilizing policy for the proposed algorithm using only measured data. Different from \cite{PoSa:22}, this method does not require the solution of linear matrix inequalities (LMIs).

In the following, Section \ref{secprel} introduces the preliminary results that are used throughout the paper. The development of the proposed efficient RL algorithm and its theoretical analysis are shown in Section \ref{secmain}. Section \ref{secprac} analyses the computational efficiency of the proposed algorithm and presents a procedure to compute the initial stabilizing gain. In Section \ref{secnumcom}, we illustrate the theoretical results with numerical examples, and Section \ref{secconc} concludes the paper.

\section{PRELIMINARIES}
\label{secprel}

In this section, we present existing results from the literature that are relevant for the remainder of this paper.

\subsection{Matrix definitions for continuous-time data}

Consider the integer $N \in \mathbb{N}$ and the positive scalar $T~\in~\mathbb{R}_+$. Let $\xi : [0, NT] \rightarrow \mathbb{R}^\sigma$, with $[0, NT] \subset \mathbb{R}$, denote a continuous-time signal of length $NT$. Using the trajectory $\xi$, we define the following matrix 
\begin{equation}
	\mathcal{H}_{T}(\xi(t)) := \left[ \begin{array}{cccc}
		\xi(t) & \xi(t+T) & \cdots & \xi(t+(N-1)T) \end{array} \right]
	\label{hankrow}
\end{equation}
for $0 \leq t \leq T$. Notice that (\ref{hankrow}) is a time-varying matrix defined on the interval $t \in [0,T]$.

Now, consider the following CT-LTI system
\begin{equation}
	\dot x(t) = Ax(t) + Bu(t),
	\label{linsys}
\end{equation}
where $x \in \mathbb{R}^n$ and $u \in \mathbb{R}^m$ are the state and input vectors of the system, respectively. The pair $(A,B)$ is assumed to be controllable throughout the paper.

Suppose that the input signal $u : [0, NT] \rightarrow \mathbb{R}^m$ is applied to (\ref{linsys}), and the resulting state trajectory $x : [0, NT] \rightarrow \mathbb{R}^n$ is collected. From (\ref{linsys}) and the definition in (\ref{hankrow}), we can write
\begin{equation*}
	\mathcal{H}_{T}(\dot x(t)) = A \mathcal{H}_{T}(x(t)) + B \mathcal{H}_{T}(u(t)).
\end{equation*}

Since it is unusual to have the state derivative $\dot x$ available as a measurement, integrate the expression above to obtain
\begin{multline*}
	\mathcal{H}_{T}(x(T)) - \mathcal{H}_{T}(x(0)) \\
	= A  \int_0^T \mathcal{H}_{T}(x(\tau)) d\tau + B \int_0^T \mathcal{H}_{T}(u(\tau)) d\tau.
\end{multline*}
For convenience of notation, define the matrices
\begin{equation}
	\tilde X = \mathcal{H}_{T}(x(T)) - \mathcal{H}_{T}(x(0)),
	\label{xmats}
\end{equation}
\begin{equation*}
	X = \int_0^T \mathcal{H}_{T}(x(\tau)) d\tau, \quad U  = \int_0^T \mathcal{H}_{T}(u(\tau)) d\tau.
\end{equation*}
Notice that the matrix $X$ (and similarly $U$) only requires the computation of integrals of the form ${ \int_0^T x(\tau + jT) d\tau }$, $j=0,\ldots,N-1$. This is simpler than the integrals computed in the existing RL literature \cite{JiJi:12,LePaCh:12,PoSa:22}.

By definition, the following expression holds
\begin{equation}
	\tilde X = A X + B U.
	\label{xtilde}
\end{equation}

\subsection{Persistence of excitation for discrete-time systems}

Define the integer constants $L,N \in \mathbb{N}$. The Hankel matrix of depth $L$ of a discrete-time sequence $\{ \mu_k \}_{k=0}^{N-1} = \{\mu_0, \, \mu_1,\, \ldots,\, \mu_{N-1} \}$, $\mu_k \in \mathbb{R}^m$, is defined as
\begin{equation*}
	H_L(\mu) := \left[ \begin{array}{cccc}
		\mu_0 & \mu_1 & \cdots & \mu_{N-L} \\
		\mu_1 & \mu_2 & \cdots & \mu_{N-L+1} \\
		\vdots & \vdots & \ddots & \vdots \\
		\mu_{L-1} & \mu_{L} & \cdots & \mu_{N-1}
	\end{array} \right].
\end{equation*}

In \cite{WillemsRapMarDe2005}, the following definition of a PE input for discrete-time systems is made.

\begin{defn}
	\label{defdtpe}
	The discrete sequence $\{ \mu_k \}_{k=0}^{N-1}$, $\mu_k \in \mathbb{R}^m$, is said to be persistently exciting of order $L$ if its Hankel matrix of depth $L$ has full row rank, i.e., 
	\begin{equation}
		\text{rank} ( H_L(\mu) ) = m L.
		\label{dtpe}
	\end{equation}
\end{defn}

It is important to highlight the fact that Definition \ref{defdtpe} provides a condition that enables a straightforward design of a PE input and that is easy to verify for any discrete sequence.

\begin{rem}
	\label{remdtpe}
	A necessary condition for (\ref{dtpe}) to hold is that $N \geq (m+1)L - 1$. This provides a minimum length for a PE input sequence.
\end{rem}

\subsection{Persistence of excitation for continuous-time systems}
\label{secpect}

It is shown in \cite{LoMu:22} that a piecewise constant input designed by exploiting Definition \ref{defdtpe} is persistently exciting for the continuous-time system (\ref{linsys}). This class of inputs is formally described in the following definition. 

\begin{defn}[Piecewise constant PE input]
	\label{defctpe}
	Consider a time interval $T > 0$ such that
	\begin{equation}
		T \neq \frac{2 \pi k}{| \mathcal{I}_m(\lambda_i - \lambda_j) |}, \qquad \forall k \in \mathbb{Z}.
		\label{assut}
	\end{equation}
	where $\lambda_i$ and $\lambda_j$ are any two eigenvalues of matrix $A$ in (\ref{linsys}), and $\mathcal{I}_m(\cdot)$ is the imaginary part of a complex number. A piecewise constant persistently exciting (PCPE) input of order $L$ for continuous-time systems is defined as ${u(t + iT) = \mu_i}$ for all $0 \leq t < T$, $i=0,\ldots,N-1$, where $\{ \mu_i \}_{i=0}^{N-1}$ is a sequence of constant vectors $\mu_i \in \mathbb{R}^m$ that is persistently exciting of order $L$ in the sense of Definition \ref{defdtpe}.
\end{defn}

\begin{rem}
	Notice that the condition (\ref{assut}) is not restrictive, even with no knowledge of the system model (\ref{linsys}). This is because the values of $T$ that make this condition fail form a set of measure zero and are unlikely to be encountered in practice.
\end{rem}

When a PCPE input is applied to system (\ref{linsys}), the obtained input-state data set satisfies an important rank condition, as shown below.

\begin{lem}[\cite{LoMu:22}]
	\label{lempe}
	Consider system (\ref{linsys}), let the pair $(A,B)$ be controllable, and let $u$ be a PCPE input of order $n+1$ as defined in Definition \ref{defctpe}. Then, the rank condition
	\begin{equation}
		\text{rank} \left( \left[ \begin{array}{c}
			\mathcal{H}_{T}(x(t)) \\ \mathcal{H}_{T}(u(t))
		\end{array} \right] \right) = \text{rank} \left( \left[ \begin{array}{c}
		\mathcal{H}_{T}(x(t)) \\ H_{1}(\mu)
	\end{array} \right] \right) = m + n
		\label{racond}
	\end{equation}
	holds for all $0 \leq t \leq T$.
\end{lem}

\begin{rem}
	In \cite{LoMu:22}, the result in Lemma \ref{lempe} was presented considering persistence of excitation of any order $L$. For simplicity of notation, we presented Lemma~\ref{lempe} directly for PE inputs of order $n+1$. This is the only order of persistence of excitation used throughout the paper.
\end{rem}

\subsection{The LQR problem and Kleinman's algorithm}

For a CT-LTI system (\ref{linsys}), the infinite-horizon LQR problem concerns determining the control input $u$ that minimizes a cost function of the form
\begin{equation}
	J(x(0),u):= \int_0^\infty \left( x^\top(t) Q x(t) + u^\top(t) R u(t) \right) dt,
	\label{ctcf}
\end{equation}
where $Q \succeq 0$ and $R \succ 0$. Throughout the paper, we assume that the pair $(A,Q^{1/2})$ is observable. This, together with the assumed controllability of $(A,B)$, implies that the optimal control input is given by $u^*(x) = -K^* x$, where
\begin{equation*}
	K^* = R^{-1}B^\top P^* 
\end{equation*}
and the matrix $P^* \succ 0$ solves the algebraic Riccati equation
\begin{equation*}
	Q + P^*A + A^\top P^* - P^*BR^{-1}B^\top P^* = 0.
\end{equation*}

In \cite{Kleinman1968}, Kleinman proposed a model-based iterative algorithm to solve the LQR problem. This algorithm starts by selecting an initial stabilizing matrix $K_0$, i.e., a matrix such that $A-BK_0$ is Hurwitz stable. At every iteration $i$, the Lyapunov equation
\begin{equation}
	P_i (A-BK_i) + (A-BK_i)^\top P_i + Q + K_i^\top R K_i = 0
	\label{klpe}
\end{equation}
is solved for $P_i$. Then, a new feedback matrix is defined as
\begin{equation}
	K_{i+1} = R^{-1} B^\top P_i.
	\label{klpi}
\end{equation}
The algorithm iterates the equations (\ref{klpe}) and (\ref{klpi}) until convergence. With the main drawback of being a model-based method, Kleinman's algorithm otherwise possesses highly attractive features. Namely, at each iteration the matrix $K_{i+1}$ is stabilizing, the algorithm converges such that 
\begin{equation*}
	\lim_{i \rightarrow \infty} K_{i+1} = K^*,
\end{equation*}
and convergence occurs at a quadratic rate \cite{Kleinman1968}.

The following section presents the main developments of this paper.

\section{AN EFFICIENT DATA-BASED ALGORITHM FOR THE CT LQR PROBLEM}
\label{secmain}

In this section, we present an efficient data-based off-policy RL algorithm to determine the optimal controller that minimizes (\ref{ctcf}). We show that the proposed procedure is equivalent to Kleinman's algorithm (\ref{klpe})-(\ref{klpi}), and therefore preserves all of its theoretical properties.  For the clarity of exposition, we introduce first a model-based algorithm that is then used as the basis of our data-based method.

\subsection{A model-based algorithm}

Combining (\ref{klpe}) and (\ref{klpi}), we readily obtain the following expressions
\begin{equation*}
	P_i A - K_{i+1}^\top R K_i + A^\top P_i - K_i^\top R K_{i+1} + Q + K_i^\top R K_i = 0
\end{equation*}
and $B^\top P_i - R K_{i+1} = 0$. Therefore, the matrix equation
\begin{multline}
	\left[ \begin{array}{cc}
		A & B \\ -RK_i & -R \end{array} \right]^\top \left[ \begin{array}{c} P_i \\ K_{i+1}	\end{array} \right] \left[ \begin{array}{cc} I_n & 0 \end{array} \right] \\
	+ \left[ \begin{array}{c} I_n \\ 0 \end{array} \right] \left[ \begin{array}{c}
		P_i \\ K_{i+1}
	\end{array} \right]^\top \left[ \begin{array}{cc}
		A & B \\ -RK_i & -R \end{array} \right]  \\
	+   \left[ \begin{array}{cc}
		Q + K_i^\top R K_i & 0 \\ 0 & 0 \end{array} \right] = 0
	\label{lyapmat}
\end{multline}
holds, where $I_n$ is an $n \times n$ identity matrix and $0$ represents a matrix of zeros with appropriate dimensions. 

Denoting the fixed matrices as
\begin{gather}
	\Phi_i := \left[ \begin{array}{cc}
		A & B \\ -RK_i & -R \end{array} \right], \quad E := \left[ \begin{array}{cc}
		I_n & 0 \end{array} \right], \nonumber \\
	\bar Q_i := \left[ \begin{array}{cc}
		Q + K_i^\top R K_i & 0 \\ 0 & 0 \end{array} \right]
\label{matdef}
\end{gather}
and the unknown matrix as
\begin{equation}
	\Theta_{i+1} := \left[ \begin{array}{c}
			P_i \\ K_{i+1}
		\end{array} \right],
	\label{desth}
\end{equation}
we can write (\ref{lyapmat}) in the compact form
\begin{equation}
	\Phi_i^\top \Theta_{i+1} E + E^\top \Theta_{i+1}^\top \Phi_i + \bar Q_i = 0.
	\label{lyapmat2}
\end{equation}

The matrix $\Theta_{i+1} \in \mathbb{R}^{(n+m) \times n}$ consists of the unknown matrices in Kleinman's algorithm, $P_i$ and $K_{i+1}$. It is of our interest to design a method in which solving a matrix equation as in (\ref{lyapmat2}) for $\Theta_{i+1}$ corresponds to solving both (\ref{klpe}) and (\ref{klpi}) simultaneously. However, it can be noted that (\ref{lyapmat2}), as it is formulated, in general does not have a unique solution $\Theta_{i+1}$. To address this issue, first express the unknown submatrices of $\Theta_{i+1}$ as
\begin{equation}
	\Theta_{i+1} = \left[ \begin{array}{c}
		\Theta_{i+1}^1 \\ \Theta_{i+1}^2
	\end{array} \right],
\label{thsub}
\end{equation}
with $\Theta_{i+1}^1 \in \mathbb{R}^{n \times n}$ and $\Theta_{i+1}^2 \in \mathbb{R}^{m \times n}$. In the following lemma, we show that there exists only one matrix $\Theta_{i+1}$ that solves (\ref{lyapmat2}) such that the submatrix $\Theta_{i+1}^1$ is symmetric.

\begin{lem}
	\label{lemuniq}
	Consider the equation (\ref{lyapmat2}) with the matrices $\Phi_i$, $E$ and $\bar{Q}_i$ defined as in (\ref{matdef}). Moreover, let the matrix $K_i$ be stabilizing. Then, there exists a unique solution (\ref{thsub}) to this equation for which $\Theta_{i+1}^1 = (\Theta_{i+1}^1)^\top$. 
\end{lem} 
\begin{proof}
	Considering the partition in (\ref{thsub}), notice that (\ref{lyapmat2}) holds for any matrix $\Theta_{i+1}$ such that
	\begin{multline*}
		A^\top \Theta_{i+1}^1 - K_i^\top R \Theta_{i+1}^2 + (\Theta_{i+1}^1)^\top A - (\Theta_{i+1}^2)^\top R K_i \\
		+ Q + K_i^\top R K_i = 0,
	\end{multline*}
	and
	\begin{equation*}
		B^\top \Theta_{i+1}^1 - R \Theta_{i+1}^2 = 0.
\end{equation*}
From the second equation it is clear that $\Theta_{i+1}^2 = R^{-1} B^\top \Theta_{i+1}^1$. Substituting this and the fact that $\Theta_{i+1}^1 = (\Theta_{i+1}^1)^\top$ in the first equation, we get
\begin{equation}
	(A-BK_i)^\top \Theta_{i+1}^1 + \Theta_{i+1}^1 (A - BK_i) + Q + K_i^\top R K_i = 0.
	\label{neop}
\end{equation}
Since $K_i$ is stabilizing, we use Lyapunov arguments to conclude that $\Theta_{i+1}^1$ (and therefore also $\Theta_{i+1}^2$) is unique.
\end{proof}

Lemma \ref{lemuniq} implies that constraining the solution of (\ref{lyapmat2}) to include a symmetric submatrix $\Theta_{i+1}^1$ leads to the desired solution (\ref{desth}). The following lemma shows that we achieve this by properly modifying $\Phi_i$ in (\ref{matdef}).

\begin{lem}
	\label{lemsym}
	Consider the matrix equation 
	\begin{equation}
		(\Phi_i^-)^\top \Theta_{i+1} E + E^\top \Theta_{i+1}^\top \Phi_i^+ + \bar Q_i = 0,
		\label{mbmat}
	\end{equation}
	where 
	\begin{equation}
		\Phi_i^+ := \left[ \begin{array}{cc}
			A + I & B \\ -RK_i & -R
		\end{array} \right], \quad \Phi_i^- := \left[ \begin{array}{cc}
		A - I & B \\ -RK_i & -R
	\end{array} \right],
		\label{phipm}
	\end{equation}
	and the matrices $E$ and $\bar Q_i$ are defined as in (\ref{matdef}). Moreover, let the matrix $K_i$ be stabilizing. Then, the solution (\ref{thsub}) of (\ref{mbmat}) is unique, and $\Theta_{i+1}^1 = (\Theta_{i+1}^1)^\top$. Moreover, the solution of (\ref{mbmat}) is also a solution of (\ref{lyapmat2}). 
\end{lem} 
\begin{proof}
	First, define the matrix
	\begin{equation*}
		S = \left[ \begin{array}{cc}
			\Theta_{i+1}^1 - (\Theta_{i+1}^1)^\top & 0 \\ 0 & 0
		\end{array} \right].
	\end{equation*}
	Using this definition, it is straightforward to express (\ref{mbmat}) in terms of the matrix $\Phi_i$ in (\ref{matdef}) as
	\begin{equation*}
		\Phi_i^\top \Theta_{i+1} E + E^\top \Theta_{i+1}^\top \Phi_i + \bar Q_i = S.
	\end{equation*}
	Notice that the left-hand side of this expression is symmetric, and therefore so must be $S$. Now, $S$ is symmetric if and only if $\Theta_{i+1}^1 = (\Theta_{i+1}^1)^\top$, that is, $S = 0$. This implies both that the solution of (\ref{mbmat}) also solves (\ref{lyapmat2}) and, by Lemma \ref{lemuniq}, that this solution is unique. 
\end{proof}

\begin{rem}
	\label{remsolver}
	Equation (\ref{mbmat}) is a case of the generalized Sylvester-transpose equation, and algorithms to solve it efficiently are well known \cite{Hajarian2016,TansriChoCha2022,SongCheZha2011}.
\end{rem}

Using this result, we formulate Algorithm 1 below. As in any policy iteration procedure, Algorithm 1 is initialized with a stabilizing matrix $K_0$. Using this matrix (as well as model knowledge), (\ref{mbmat}) is solved for $\Theta_{i+1}$. Then, partitioning $\Theta_{i+1}$ as in (\ref{thsub}), a new feedback matrix is obtained as $K_{i+1}=\Theta_{i+1}^2$. 

\begin{figure}[h]
	\hrule
	{\bf Algorithm 1: Model-based RL algorithm}
	{\hrule \small
		\begin{algorithmic}[1]
			\Procedure{}{}
			\State Let $i=0$ and initialize a stabilizing feedback matrix $K_0$.
			\State Using the definitions in (\ref{matdef}) and (\ref{phipm}), solve for $\Theta_{i+1}$ from the equation 
			\begin{equation*}
				(\Phi_i^-)^\top \Theta_{i+1} E + E^\top \Theta_{i+1}^\top \Phi_i^+ + \bar Q_i = 0.
			\end{equation*}
			\State Partitioning $\Theta_{i+1}$ as in (\ref{thsub}), define
			\begin{equation*}
				K^{i+1}=\Theta_{i+1}^2.
			\end{equation*}
			\State If $\| K^{i+1} - K^i \| > \varepsilon$ for some $\varepsilon > 0$, let $i=i+1$ and go to Step 3. Otherwise, stop. 
			\EndProcedure
			\hrule
		\end{algorithmic}
	}
\end{figure}

Using the results obtained so far, we conclude that Algorithm 1 is equivalent to Kleinman's algorithm in the sense that, starting from the same initial matrix $K_0$, they provide the same updated policies $K_{i+1}$ at every iteration. This implies that Algorithm 1 preserves all the properties of Kleinman's algorithm. In the following, we use this result to design a data-based algorithm.

\subsection{The data-based algorithm}

To avoid the need for model knowledge in Algorithm 1, we collect persistently excited data from the system (\ref{linsys}) as described in Section \ref{secpect}. Using this data, we define the constant matrices $X$, $U$ and $\tilde X$ as in (\ref{xmats}).

Lemma \ref{lempe} showed that the collected data set satisfies the rank condition (\ref{racond}). In the following lemma, we extend this result to the matrices $X$ and $U$.

\begin{lem}
	\label{lemxu}
	Consider system (\ref{linsys}), let the pair $(A,B)$ be controllable, and let $u$ be a PCPE input of order $n+1$ as defined in Definition \ref{defctpe}. Using the resulting input-state data, define the matrices $X$ and $U$ as in (\ref{xmats}). Then,
	\begin{equation}
		\text{rank} \left( \left[ \begin{array}{c} X \\ U	\end{array} \right] \right) = n + m.
	\end{equation}
\end{lem}
\begin{proof}
	Notice that, since the applied input is piecewise constant, an expression for the resulting state of (\ref{linsys}) is
	\begin{equation*}
		x(t + iT) = e^{At} x(iT) + \int_0^t e^{A\tau} d\tau B \mu_i,
	\end{equation*}
	for $i=0,\ldots, N-1$ and $0 \leq t \leq T$. Thus, we can write
	\begin{multline*}
		\left[ \begin{array}{c} X \\ U \end{array} \right] = \int_0^T \left[ \begin{array}{c} \mathcal{H}_T(x(\tau)) \\ H_1(\mu) \end{array} \right] d\tau \\
		= \underbrace{\int_0^T \left[ \begin{array}{cc} e^{A\tau} & \int_0^\tau e^{As} ds B \\ 0 & I \end{array} \right] d\tau}_{W} \left[ \begin{array}{c} \mathcal{H}_T(x(0)) \\ H_1(\mu) \end{array} \right].
	\end{multline*}
	Notice that $W$ is nonsingular since the condition (\ref{assut}) holds (the fact that $\int_0^T e^{A\tau} d\tau$ is nonsingular follows from the fact that $T$ corresponds to a non-pathological sampling time \cite{Chen1999}). Moreover, by Lemma~\ref{lempe} the second matrix on the right-hand side has full row rank, completing the proof.  
\end{proof}

Define $Z = [X^\top \quad U^\top]^\top$. Since $Z$ has full row rank by Lemma \ref{lemxu}, we can select $n+m$ linearly independent columns from it. Let $z_k$ represent the $k$th column of $Z$, and let $\eta = \{ k_1,\ldots,k_{n+m} \}$ be a set of indices such that
\begin{equation}
	Z_\eta := \left[ z_{k_1} \quad \cdots \quad z_{k_{n+m}} \right]
	\label{zeta}
\end{equation}
is a nonsingular matrix. Then, $\Theta_{i+1}$ is a solution of (\ref{mbmat}) if and only if it is a solution of
\begin{equation}
	Z_\eta^\top (\Phi_i^-)^\top \Theta_{i+1} E Z_\eta + Z_\eta^\top E^\top \Theta_{i+1}^\top \Phi_i^+ Z_\eta + Z_\eta^\top \bar Q_i Z_\eta = 0.
	\label{mbmat2}
\end{equation}
From the definitions in (\ref{matdef}) and (\ref{phipm}), and using the expression (\ref{xtilde}), we have the following
\begin{equation*}
	\Phi_i^+ Z_\eta = \left[ \begin{array}{c} AX_\eta + X_\eta + BU_\eta \\ -RK_i X_\eta - R U_\eta \end{array} \right] = \left[ \begin{array}{c} \tilde X_\eta + X_\eta \\ -RK_i X_\eta - R U_\eta \end{array} \right],
\end{equation*}
\begin{equation*}
	\Phi_i^- Z_\eta = \left[ \begin{array}{c} AX_\eta - X_\eta + BU_\eta \\ -RK_i X_\eta - R U_\eta \end{array} \right] = \left[ \begin{array}{c} \tilde X_\eta - X_\eta \\ -RK_i X_\eta - R U_\eta \end{array} \right],
\end{equation*}
$Z_\eta^\top \bar Q_i Z_\eta = X_\eta^\top (Q + K_i^\top R K_i) X_\eta$ and $EZ_\eta = X_\eta$, where the subindex $\eta$ represents a matrix constructed using the columns specified by the set $\eta$ from the corresponding original matrix. Substituting in (\ref{mbmat2}), we obtain
\begin{equation}
	(Y_i^-)^\top \Theta_{i+1} X_\eta + X_\eta^\top \Theta_{i+1}^\top Y_i^+ + X_\eta^\top (Q + K_i^\top R K_i) X_\eta = 0.
	\label{dbmatte}
\end{equation}
where
\begin{equation}
	\begin{aligned}
	Y_i^- & := \left[ \begin{array}{c} \tilde X_\eta - X_\eta \\ -RK_i X_\eta - R U_\eta \end{array} \right], \\
	Y_i^+ & := \left[ \begin{array}{c} \tilde X_\eta + X_\eta \\ -RK_i X_\eta - R U_\eta \end{array} \right].
	\end{aligned}
	\label{xpm}
\end{equation}

Now, (\ref{dbmatte}) is a data-based equation that does not require any knowledge about the system model. Algorithm 2 uses this expression to solve the LQR problem. For convenience, for Algorithm 2 we define
\begin{equation}
	Q_i := X_\eta^\top (Q + K_i^\top R K_i) X_\eta.
	\label{qi}
\end{equation}

\begin{figure}[h]
	\hrule
	{\bf Algorithm 2: Data-based RL algorithm}
	{\hrule \small
		\begin{algorithmic}[1]
			\Procedure{}{}
			\State Select $N \geq (n+1)m+n$ and $T > 0$, apply a PCPE input of order $n+1$ to (\ref{linsys}) and collect an $NT$-long input-state trajectory.
			\State Compute the matrices $X$, $U$, and $\tilde X$ as in (\ref{xmats}).
			\State Select a set of indices $\eta = \{ k_1,\ldots,k_{n+m} \}$ such that $[X_\eta^\top \quad U_\eta^\top ]^\top$ is nonsingular.
			\State Let $i=0$ and initialize a stabilizing feedback matrix $K_0$.
			\State Define the matrices $Y_i^+$, $Y_i^-$ and $Q_i$ as in (\ref{xpm})-(\ref{qi}), and solve for $\Theta_{i+1}$ from the equation 
			\begin{equation}
				(Y_i^-)^\top \Theta_{i+1} X_\eta + X_\eta^\top \Theta_{i+1}^\top Y_i^+ + Q_i = 0.
				\label{dbmat}
			\end{equation}
			\State Partitioning $\Theta_{i+1}$ as in (\ref{thsub}), define
			\begin{equation*}
				K^{i+1}=\Theta_{i+1}^2.
			\end{equation*}
			\State If $\| K^{i+1} - K^i \| > \varepsilon$ for some $\varepsilon > 0$, let $i=i+1$ and go to Step 6. Otherwise, stop. 
			\EndProcedure
			\hrule
		\end{algorithmic}
	}
\end{figure}

The following theorem states the main properties of this algorithm.

\begin{thm}
	Consider the CT-LTI system (\ref{linsys}), and the partitioning (\ref{thsub}) of $\Theta_{i+1}$. Every iteration of Algorithm 2 has the following properties: ($i$) the solution $\Theta_{i+1}$ of (\ref{dbmat}) exists and is unique; ($ii$) the gain $K_{i+1}$ is stabilizing; and ($iii$) $\Theta_{i}^1 \succeq \Theta_{i+1}^1 \succeq P^*$. Moreover,
	\begin{equation*}
		\lim_{i \rightarrow \infty} K_i = K^*
	\end{equation*}
	and the rate of convergence of the algorithm is quadratic.
\end{thm}
\begin{proof}
	The proof is obtained by showing that Algorithm~2 is equivalent to Kleinman's algorithm at every iteration. First, notice that by Lemma $\ref{lemxu}$, the matrix $[ X^\top \quad U^\top ]^\top$ has full row rank and, therefore, a nonsingular matrix $[ X_\eta^\top \quad U_\eta^\top ]^\top$ can always be constructed. This means that (\ref{dbmat}) is equivalent to (\ref{mbmat}). Now, noting that $K_0$ is stabilizing, use an induction argument to assume that $K_i$ is stabilizing. Lemma \ref{lemsym} shows the existence and uniqueness of $\Theta_{i+1}$ from (\ref{mbmat}). Moreover, the expression (\ref{neop}) in the proof of Lemma \ref{lemuniq} shows that $\Theta_{i+1}^1 = P_i$, where $P_i$ is the solution of the Lyapunov equation (\ref{klpe}). Also in the proof of Lemma \ref{lemuniq} it was shown that $\Theta_{i+1}^2 = R^{-1}B^\top \Theta_{i+1}^1$, which now corresponds to Kleinman's updated gain (\ref{klpi}). Therefore, Algorithm 2 is equivalent to Kleinman's algorithm and shares all of its properties \cite{Kleinman1968}.
\end{proof}

Algorithm 2 is a purely data-based, off-policy method to solve the continuous-time LQR problem. Using Definition \ref{defctpe}, we are able to guarantee the existence of a solution $\Theta_{i+1}$ of (\ref{dbmat}) at every iteration for data trajectories of fixed length. This contrasts with the methods in the literature that must keep collecting data until a matrix gets full rank, such as, e.g., \cite{PoSa:22,JiJi:12}. Moreover, we avoid the use of the Kronecker product and its resulting large matrices in Algorithm 2. As stated in Remark \ref{remsolver}, methods to efficiently solve a Sylvester-transpose equation as in (\ref{dbmat}) are well known.

\begin{rem}
	Step 4 of Algorithm 2 instructs to select $n+m$ linearly independent columns of $[X^\top \quad U^\top]^\top$. This step is performed in benefit of efficiency, as it decreases the size of the matrices in (\ref{dbmat}). However, since $[X^\top \quad U^\top]^\top$ has full row rank, skipping this step in Algorithm 2 and using the complete data matrices instead does not affect the result at each iteration.  
\end{rem}

\section{PRACTICAL CONSIDERATIONS}
\label{secprac}

\subsection{Efficiency analysis of Algorithm 2}

In this subsection, we analyze the theoretical computational complexity of Algorithm 2. Moreover, we compare this complexity with that of the algorithm proposed in \cite{JiJi:12}. This is because \cite{JiJi:12} is also an off-policy data-based method that shares many of the characteristics of Algorithm 2. 

The most expensive steps in Algorithm 2 are obtaining the solution of (\ref{dbmat}) and selecting $n+m$ linearly independent vectors from $[X^\top \quad U^\top]^\top$. Methods to solve the Sylvester-transpose equation (\ref{dbmat}) with a complexity of $\mathcal{O}((n+m)^3)$ are known \cite{TansriChoCha2022}. The selection of linearly independent vectors can be performed using a simple procedure like Gaussian elimination to transform the matrix of interest into row echelon form. This method has a complexity of $\mathcal{O}((n+m)^2N)$ operations \cite{Strang:06}. This step, however, only needs to be performed once in Algorithm 2 (in Step 4). Thus, we conclude that Algorithm 2 requires once $\mathcal{O}((n+m)^2N)$ and then in each iteration $\mathcal{O}((n+m)^3)$ floating point operations.

The algorithm in \cite{JiJi:12} was also shown to be equivalent to Kleinman's algorithm at every iteration. However, their method uses a Kronecker product formulation that yields matrices of large dimensions. Let $N_\otimes$ be the amount of data samples used in \cite{JiJi:12}. Then, the most expensive step at each iteration of their algorithm is the product of a matrix with dimensions $(\frac{1}{2} n (n+1) + mn) \times N_\otimes$ times its transpose. This product, and hence each iteration of the algorithm, requires $\mathcal{O}((\frac{1}{2} n (n+1) + mn)^2 N_\otimes)$ floating point operations \cite{Hunger2007}. Clearly, as the dimension of the system increases, the difference in performance of both algorithms becomes more significant. Moreover, we notice from \cite{JiJi:12} that the amount of collected data must satisfy $N_\otimes \geq \frac{1}{2} n (n+1) + mn$ for the algorithm to yield a unique solution at every iteration. Compare this with the bound $N \geq (n+1) m + n$ in Algorithm~2. In Section \ref{secnumcom}, we test this theoretical comparison using numerical examples.

\subsection{An initial stabilizing policy}

In \cite[Remark 2]{DePersisTes2020}, a procedure to design a stabilizing controller for continuous-time systems using only measured data was described. This method is based on the solution of a linear matrix inequality (LMI). The authors in \cite{PoSa:22} proposed to use a similar LMI-based procedure to determine the initial stabilizing gain for a Q-learning algorithm. Since one of the goals in this paper is computational efficiency, we would like to avoid the computationally expensive step of solving an LMI. In this subsection, we present an alternative method to determine the initial stabilizing matrix $K_0$ for Algorithm~2. The following development follows closely a procedure proposed in \cite[Section IV]{vanWaardeEisTreCam2020} for discrete-time systems.

Let $F$ be the Moore-Penrose pseudoinverse of the matrix $X$ in (\ref{xmats}). Since $X$ has full row rank (see Lemma \ref{lemxu}), $F$ is a right inverse of $X$. Furthermore, let $G$ be a basis for the null space of $X$, such that $X (F - G \bar K) = I$ for any matrix $\bar K$ of appropriate dimensions. Using the matrices $F$, $G$ and $U$ from (\ref{xmats}), we propose to compute the initial stabilizing gain $K_0$ for Algorithm 2 as
\begin{equation}
	K_0 = - U (F - G \bar K)
	\label{inik}
\end{equation}
where $\bar K$ is a matrix to be determined.

From (\ref{xtilde}) and (\ref{inik}), notice that
\begin{align*}
	\tilde X (F - G \bar K) & = \left[ A \quad B \right] \left[ \begin{array}{c} X \\ U \end{array} \right] (F - G \bar K) \\
	& = \left[ A \quad B \right] \left[ \begin{array}{c} I \\ -K_0 \end{array} \right]
\end{align*}
Therefore, by designing the poles of the matrix $\tilde X (F - G \bar K)$, we also set the poles of $A - BK_0$. Since $(A,B)$ is controllable and hence the poles of $A - BK_0$ can be assigned arbitrarily, also the poles of $\tilde X (F - G \bar K)$ can be placed arbitrarily by a suitable choice of $\bar K$. Moreover, since $\tilde X$, $F$ and $G$ are matrices obtained from data, we can operate with them without any need of model knowledge. This procedure is summarized in the following theorem. The proof of this theorem is straightforward considering the procedure described in this subsection and is hence omitted.

\begin{thm}
	\label{thinik}
	Let the matrices $\tilde X$, $X$ and $U$ be defined as in (\ref{xmats}) using data collected from (\ref{linsys}) during the application of a PCPE input of order $n+1$. Define $F$ as the Moore-Penrose pseudoinverse of $X$ and $G$ as a basis for the null space of $X$. Moreover, define the \emph{virtual} system matrices $\bar A = \tilde X F$ and $\bar B = \tilde X G$. Using pole-placement methods, determine a matrix $\bar K$ such that $\bar A - \bar B \bar K$ is Hurwitz. Then, the matrix $K_0$ defined by (\ref{inik}) is stabilizing for system (\ref{linsys}).
\end{thm}

\begin{rem}
	Notice that the matrices $\bar A = \tilde X F$ and $\bar B = \tilde X G$ in Theorem \ref{thinik} do not correspond to the actual system matrices $A$ and $B$. In fact, $B$ and $\bar B$ in general do not have the same dimensions. No model identification is performed in the proposed procedure.
\end{rem}

\addtolength{\textheight}{-2.5cm}   %

\section{NUMERICAL EXPERIMENTS}
\label{secnumcom}

In this section, we compare in simulation the efficiency of the proposed Algorithm 2 with that of the algorithm presented in \cite{JiJi:12}. As described above, these algorithms have the same characteristics: they are data-based off-policy methods that are equivalent to Kleinman's algorithm at every iteration.

 To compare the efficiency of both algorithms, several simulations are performed for different, randomly generated linear systems (\ref{linsys}). In particular, 100 different linear systems are generated using the command \emph{rss} in Matlab, and both algorithms are applied to each of them. The system dimensions considered for each set of 100 experiments are $n=2,3,5$ and $7$. In every case, we consider single input systems ($m=1$), and we define the cost function (\ref{ctcf}) with $Q=I$ and $R=2$.
 
 Each implementation of Algorithm 2 had the following characteristics. A PCPE input as in Definition \ref{defctpe} was used to collect data from the system. A sample of data was collected every $10^{-4}$ time units. We considered a time interval of $T=0.2$, and we collected data for a total of $NT$ time units, with $N = (n+1)m+n$. The method described in \cite{Hajarian2016} was used to solve the Sylvester-transpose equation (\ref{dbmat}) at every iteration. 
 
 For the implementation of the Kronecker product-based method in \cite{JiJi:12}, we followed the same simulation characteristics described in the simulation section of that paper. The only exception is in the amount of data collected, which was reduced for small system dimensions in order to make a fairer comparison. 
 
 Finally, notice that the command \emph{rss} in Matlab yields stable systems. Thus, an initial stabilizing matrix of $K_0 = 0$ was used for all experiments and both algorithms. The simulations were performed using Matlab R2020b on an Intel i7-10875H (2.30 GHz) with 16 GB of memory. 
 
 The results of our simulations are displayed in Table \ref{tab1}. In this table, we refer to Algorithm 2, which is based on the solution of a Sylvester-transpose equation, as `SYL'. The algorithm in \cite{JiJi:12} that is based on the use of the Kronecker product is denoted as `KRO'. To compare the computational efficiency of the methods, we present the average time that it takes the algorithms to complete 10 iterations. Due to their quadratic rate of convergence, 10 iterations yield a very accurate result of the optimal control gain for both algorithms. In the table we can observe a confirmation of our theoretical analysis regarding the improved performance of Algorithm 2.
 
\begin{table}
	\centering
	\begin{tabular}{|c|c c|} 
		\hline
		\multirow{2}{1.5cm}{\centering Dimension $n$} & \multicolumn{2}{c|}{Average time (sec)} \\
		& SYL & KRO \\[0.5ex]
		\hline
		$2$ & $0.0099$ & $0.0953$ \\ 
		$3$ & $0.0144 $ & $0.1719$ \\
		$5$ & $0.0287$ & $0.4317$ \\
		$7$ & $0.1046$ & $1.8021$ \\ [1ex] 
		\hline
	\end{tabular}
	\vspace{4pt}
	\caption{Run-time comparison between Algorithm 2 (SYL) and the algorithm in \cite{JiJi:12} (KRO).}
	\label{tab1}
\end{table}

During the execution of these experiments, we noted some issues in the performance of both methods when applied to systems of large dimensions. First, Algorithm 2 requires the application of a solver from the literature to solve (\ref{dbmat}). We found that, if the data matrix $Z_\eta$ in Algorithm 2 has a large condition number, the solvers considered often failed to provide the correct result. To address this problem, methods to construct a matrix $Z_\eta$ with low condition number from a larger matrix $Z$ could be considered. Regarding the algorithm in \cite{JiJi:12}, determining a proper input in order to satisfy the required persistence of excitation condition for the collected data (compare the discussion in the Introduction) becomes ever more difficult as the dimension of the system increases. In this case, it is uncertain how to solve this issue.

\section{CONCLUSIONS}
\label{secconc}

In this paper, a computationally efficient algorithm was proposed to solve the continuous-time LQR problem. The proposed algorithm is equivalent to Kleinman's method, it does not require any knowledge from the system model and it requires collecting data from the system only once. We presented a persistently exciting input that guarantees that the matrix equation (\ref{dbmat}) in our algorithm has a unique solution at every iteration. Finally, we showed a method to determine an initial stabilizing feedback matrix using only measured data and that does not require to solve LMIs. Simulation results show that our algorithm significantly improves the performance of an algorithm with similar properties in the literature.


\bibliographystyle{IEEEtran}
\bibliography{IEEEabrv,ctlqr_refs}

\end{document}